# Homogeneous pressure decay in multi-chamber vacuum system with finite conductance

Fernando F. Dall'Agnol[1*] Felipe Vieira[1] and Francisco T. Degasperi[2].

[1]Santa Catarina Federal University (UFSC), R. João Pessoa 2514, Velha, Blumenau – SC, Brazil, 89036-004.

[2]Faculdade de Tecnologia de São Paulo, Av. Tiradentes, 615 - Bom Retiro, São Paulo – SP, Brazil, 01124-060.

*Corresponding author: fernando.dallagnol@ufsc.br

## Abstract

We derived the geometrical parameters on the tube connections that homogenize the pressure drop in a multi-chamber vacuum system, where each chamber has a distinct volume and all are connected to the same vacuum pump. We start deriving the pressure drop in a single chamber for a tube with finite conductance. Next, we derive a solution that provides the radius and length for each tube connection between the chambers and the pump that homogenize the pressure drop in all chambers.

## List of Symbols

The table below defines the variables used in this work.

| | Definition | Unit SI |
|---|---|---|
| $V$ | Total volume of the vacuum chambers. | $m^3$ |
| $S$ | Total volumetric rate from the vacuum pump (Pumping speed). | $m^3/s$ |
| $R$ | Radius of the tube in a single chamber system. | m |
| $L$ | Length of the tube in a single chamber system. | m |
| $q$ | Volumetric rate exiting the chamber. | $m^3/s$ |
| $q_2$ | Volumetric rate into the vacuum pump. | $m^3/s$ |
| $K$ | Volumetric rate per unit of pressure difference | $m^3s^{-1}Pa^{-1}$ |
| $-dm$ | Element of mass removed from the chamber. | kg |
| $\rho$ | Density of the gas in the chamber. | $kg/m^3$ |
| $\rho_2$ | Density of the gas at the vacuum pump. | $kg/m^3$ |
| $p$ | Pressure of the gas in the chamber. | Pa |
| $p_0$ | Initial pressure in the chamber. | Pa |
| $p_2$ | Pressure of the gas at the pump. | Pa |
| $p_R$ | Reference pressure defined in the text. | Pa |
| $\Delta p$ | Pressure difference between the ends of the tube. | Pa |
| $\mu$ | Dynamic viscosity of the gas. | Pa.s |
| $V_i$ | Volume of the $i^{th}$ chamber in a multi chamber system. | $m^3$ |
| $S_i$ | Pump speed at the $i^{th}$ tube in a multi chamber system. | $m^3/s$ |
| $R_i$ | Radius of the $i^{th}$ tube in a multi chamber system. | m |
| $L_i$ | Length of the $i^{th}$ tube in a multi chamber system. | m |
| $K_i$ | Volumetric rate per unit of pressure difference from the $i^{th}$ tube. | $m^3s^{-1}Pa^{-1}$ |
| $t$ | Time. | s |
| $\Delta t$ | Evacuation time interval | s |
| $\Delta t_\infty$ | Evacuation time interval for $K \to \infty$. | s |
| $p_f$ | Final pressure after $\Delta t$. | Pa |

## 1 Introduction

There are vacuum systems consisting of multiple chambers, which can conveniently be attached to the same vacuum pump in benefit of the project's simplicity, easy installation, easy operation, and maintenance. Some of these applications requires only rough vacuum, for example, in smelteries and laboratory vacuum lines, medicaments, petrochemistry [1]. All these industries can benefit from a single large pump. Under rough vacuum, the gas obeys the Navier-Stokes equation, which has analytical solution for tubes of circular cross section. In this work, we specify the characteristics of the tubes connecting the chambers to the pump that homogenize the evacuation in all chambers, i.e., the pressure shall drop uniformly in all chamber, as well as, the evacuation time is also the same for all chambers.

We start our analysis deriving the volumetric rate (in $m^3/s$) from a circular long tube for a compressible fluid. This derivation is fairly straightforward; however, it is very difficult to find in the literature. Possibly, it was never published. Next, we determine the conditions on the tubes that homogenize the pressure drop. This article is aimed at providing a solution to a problem in Vacuum Technology, however, we opted to use the jargon of the Fluid Dynamics that we use more often in related works. With this, we also aim to promote the application of this work in computer fluid dynamics and any computer code that uses mostly the Fluid Dynamics nomenclature.

## 2 Initial considerations

To obtain a manageable solution to our problem we must accept a few reasonable approximations. We assumed that:

i. The tubes cross section are perfect circles, which have the simplest solution for the volumetric rate.

ii. The pressure is modeled within the rough vacuum approximation, >100 Pa (1 mbar), where the fluid can be considered a continuous medium that obeys the Navier-Stokes equation.

iii. The gas in the chamber is considered ideal, so Clapeyron's equation $pV=nRT$ is valid.

iv. We ignore the transient while the fluid is accelerated from rest, assuming that this time interval is negligibly small compared to the time scale of the pressure drop from ~100 kPa to 100 Pa.

v. The mass stored in the tube does not vary appreciable during the evacuation time. Hence, the mass rates at both ends are considered equal.

vi. The whole system is isothermal. We neglect temperature variations when the fluid undergoes expansion due to the pressure drop. This approximation is reasonable if the pressure drop is slow.

vii. The conductance of the orifices that constitutes the ends of the tube are neglected.

# 3 Single chamber system

In this section, we evaluate the pressure drop dependency with all physical and geometrical parameters. Consider the vacuum chamber linked to the vacuum pump through a long thin tube as in Figure 1. Vacuum pumps are usually placed close to the chamber such that the length of the tube does not hinder the pressure drop, however, in our study the long tube is a necessity to permit a single pump to connect multiple chambers. Hence the pressure drop due to the viscosity has to be taken into account.

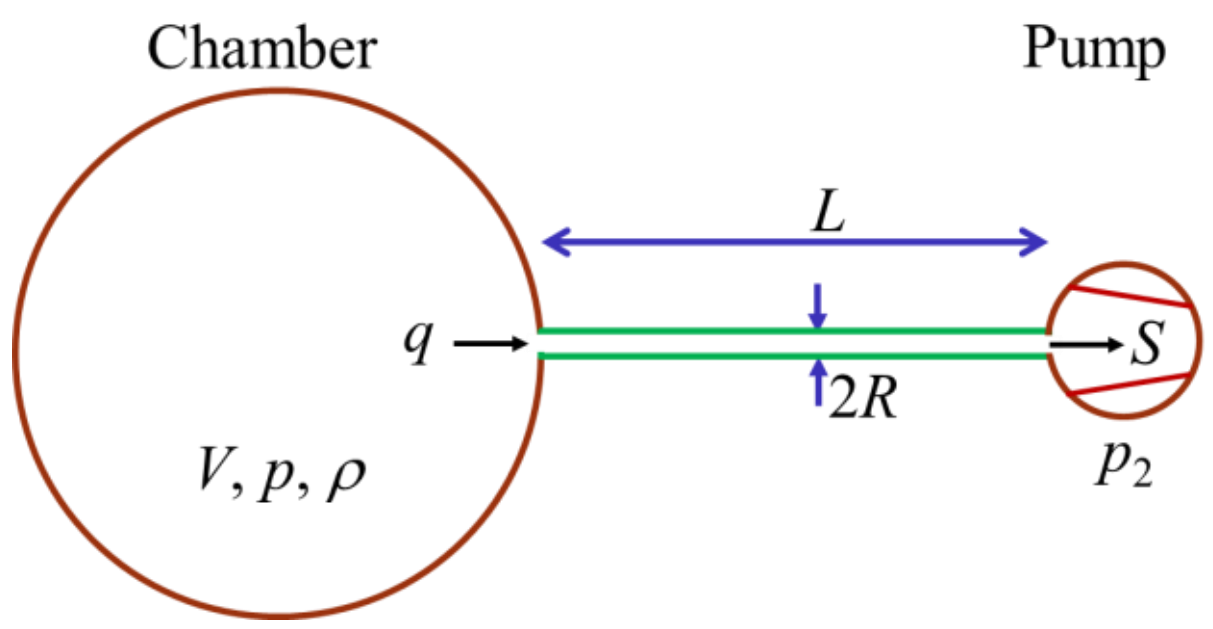

Figure 1: Single chamber vacuum system showing the geometrical parameters considered in our analysis.

## 3.1 Basic equations

The volumetric rate, at the right-hand side is, by definition, the pump rate (or pump speed) *S*. It is also well known that the volumetric rate for a compressible gas in a circular tube is given by Hagen-Poiseuille equation [2] (we endorse the derivation from Wikepedia in the footnote *):

$$S = K\frac{(p^2 - p_2^2)}{2p_2}, \tag{1}$$

where $K=\pi R^4/(8\mu L)$, $p$ is the pressure in the vacuum chamber and $p_2$ is the pressure at the right-hand end of the tube. The $\mu$ is the dynamic viscosity, $R$ and $L$ are the radius and length of the tube. We can rewrite (1) as a quadratic equation on $p_2$:

$$p_2^2 + 2\frac{S}{K}p_2 - p^2 = 0, \tag{2}$$

with positive solution

$$p_2 = \sqrt{p^2 + p_R^2} - p_R, \tag{3}$$

* https://en.wikipedia.org/wiki/Hagen-Poiseuille_equation.

where $p_R = S/K$ is a *reference pressure* defined here to simplify the equations. The $p_R$ has a physical interpretation; it is the pressure difference, which would generate a flow that equals the pump rate in an incompressible fluid.

Next, we delve in the algebra to obtain the evolution of the pressure.

## 3.2 Dynamics of the pressure in the chamber

The conservation of mass, plus the assumption (iv) from Section 2, requires that the mass rate on both ends of the tube is the same:

$$\left.\frac{dm}{dt}\right|_1 = \left.\frac{dm}{dt}\right|_2 , \tag{4}$$

hence,

$$V\frac{d\rho}{dt} = -\rho_2 S , \tag{5}$$

where $\rho$ and $\rho_2$ are the mass densities at each end. In ideal gases, the densities are proportional to the pressure, therefore $\rho$ and $\rho_2$ can be respectively replaced by $p$ and $p_2$, where the latter is given by (3), and we obtain:

$$V\frac{dp}{dt} = -\left(\sqrt{p^2 + p_R^2} - p_R\right)S . \tag{6}$$

This is a relatively simple, although non-linear, first order differential equation. The variables can be separated and integrated

$$\int_{p_0}^{p} \frac{1}{\sqrt{p^2 + p_R^2} - p_R} dp = -\frac{S}{V} t , \tag{7}$$

resulting in

$$\ln\left(\frac{p+\sqrt{p^2+p_R^2}}{p_0+\sqrt{p_0^2+p_R^2}}\right)+\frac{p_R+\sqrt{p^2+p_R^2}}{p}-\frac{p_R+\sqrt{p_0^2+p_R^2}}{p_0}=-\frac{S}{V}t \quad . \tag{8}$$

Or alternatively, by exponentializing both members, we ca write (8) as:

$$\left(p+\sqrt{p^2+p_R^2}\right)\exp\left(-\frac{p_R+\sqrt{p^2+p_R^2}}{p}\right)=\left(p_0+\sqrt{p_0^2+p_R^2}\right)\exp\left(-\frac{p_R+\sqrt{p_0^2+p_R^2}}{p_0}\right)\exp\left(-\frac{S}{V}t\right) . \tag{9}$$

The solution in the form of eq. (8) have been previously derived in the literature, particularly in the work of A. Roth [3], and before him, Delafosse and Mongodin [4]. However, these works are old and not redly available, so we reproduced them here as well.

Note that if the tube conductance is large, i.e., $K\rightarrow\infty$, then $p_R\rightarrow 0$. Consequently, (9) reduce to the exponential decay, as expected:

$$p = p_0 \exp\left(-\frac{S}{V}t\right) . \tag{10}$$

On the other hand, if the conductance is severely restricted ($K\rightarrow 0$), then (9) reduces to:

$$p=\frac{p_0}{1+\frac{p_0}{2p_R}\frac{S}{V}t} . \tag{11}$$

Surprisingly, the pressure does not decay exponentially at low conductivities; it decays hyperbolically instead.

For intermediate values $0 < K < \infty$, (9) is transcendental, so the pressure cannot be isolated and numerical techniques must be employed to determine $p(t)$ (Newton’s method should be good enough). Once determined, we can insert $p(t)$ back in all previous physical quantities, $\{q, q_2, \rho, \rho_2, p_2\}(t)$ to depict the time evolution of the system entirely.

### 3.3 Evacuation time

We want to compare the evacuation time interval from $p_0$ to final pressure $p_f$ considering infinite and finite conductivities. Let $\Delta t_\infty$ be the evacuation time interval when $K \rightarrow \infty$, or equivalently, $p_R \rightarrow 0$. In this case

$$\Delta t_\infty = \frac{V}{S} \ln\left(\frac{p_0}{p_f}\right). \quad (12)$$

Otherwise, if $p_R \neq 0$, it follows from isolating $t$ in (9):

$$\Delta t = \frac{V}{S}\left[\ln\left(\frac{p_0 + \sqrt{p_0^2 + p_R^2}}{p_f + \sqrt{p_f^2 + p_R^2}}\right) + \frac{p_R + \sqrt{p_f^2 + p_R^2}}{p_f} - \frac{p_R + \sqrt{p_0^2 + p_R^2}}{p_0}\right] \quad (13)$$

### 3.4 Evacuation time exemplification

Figure 2 shows typical pressure drops for several values of $K$. Parameters used were, $\mu = 2\times10^{-5}$ P, $V$=0.02 m$^3$, $S=10^{-4}$ m$^3$/s, $L$=1 m, $p_0$=100 kPa. The curve for $K=1.96\times10^{-8}$ m$^3$s$^{-1}$Pa$^{-1}$ ($R$=1 mm) takes $\Delta t$=20900 s, or 15 $\Delta t_\infty$ to drop to 100 Pa, where $\Delta t_\infty$=1382 s. However, the curve for $K=4.93\times10^{-6}$ m$^3$s$^{-1}$Pa$^{-1}$ ($R$=4 mm) takes just $\Delta t$=1423 s, or 1.03 $\Delta t_\infty$.

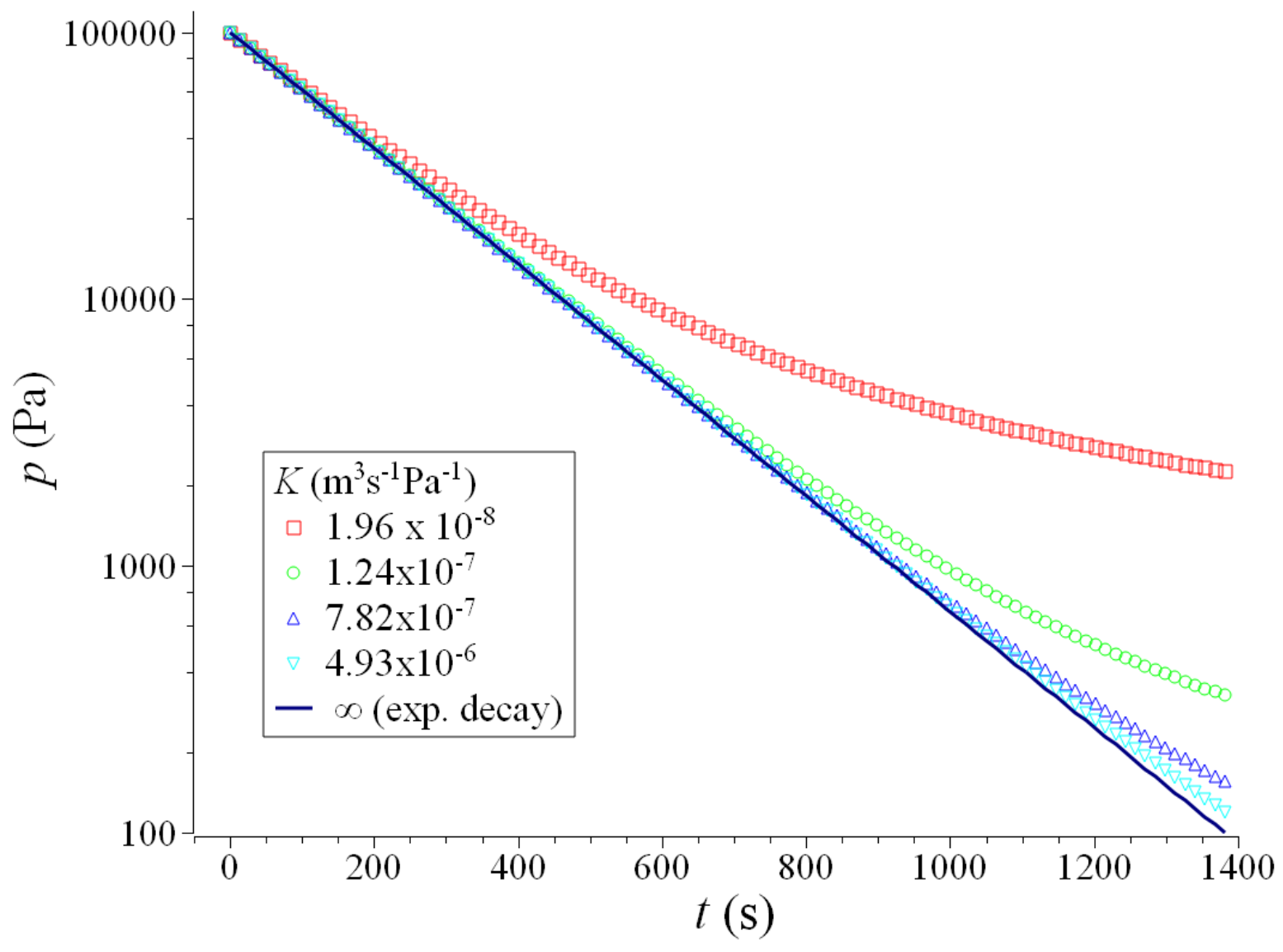

Figure 2: Pressure chamber for various values of $K=\pi R^4/(8\mu L)$. As $K$ increases, the pressure decay tends to the exponential decay.

# 4 Multi chamber system

Given a distribution of $n$ chambers and the position of the pump, we aim to answer what are the set $\{K_i\}$ that homogenize the pressure drop in all chambers, considering that all tubes are in parallel.

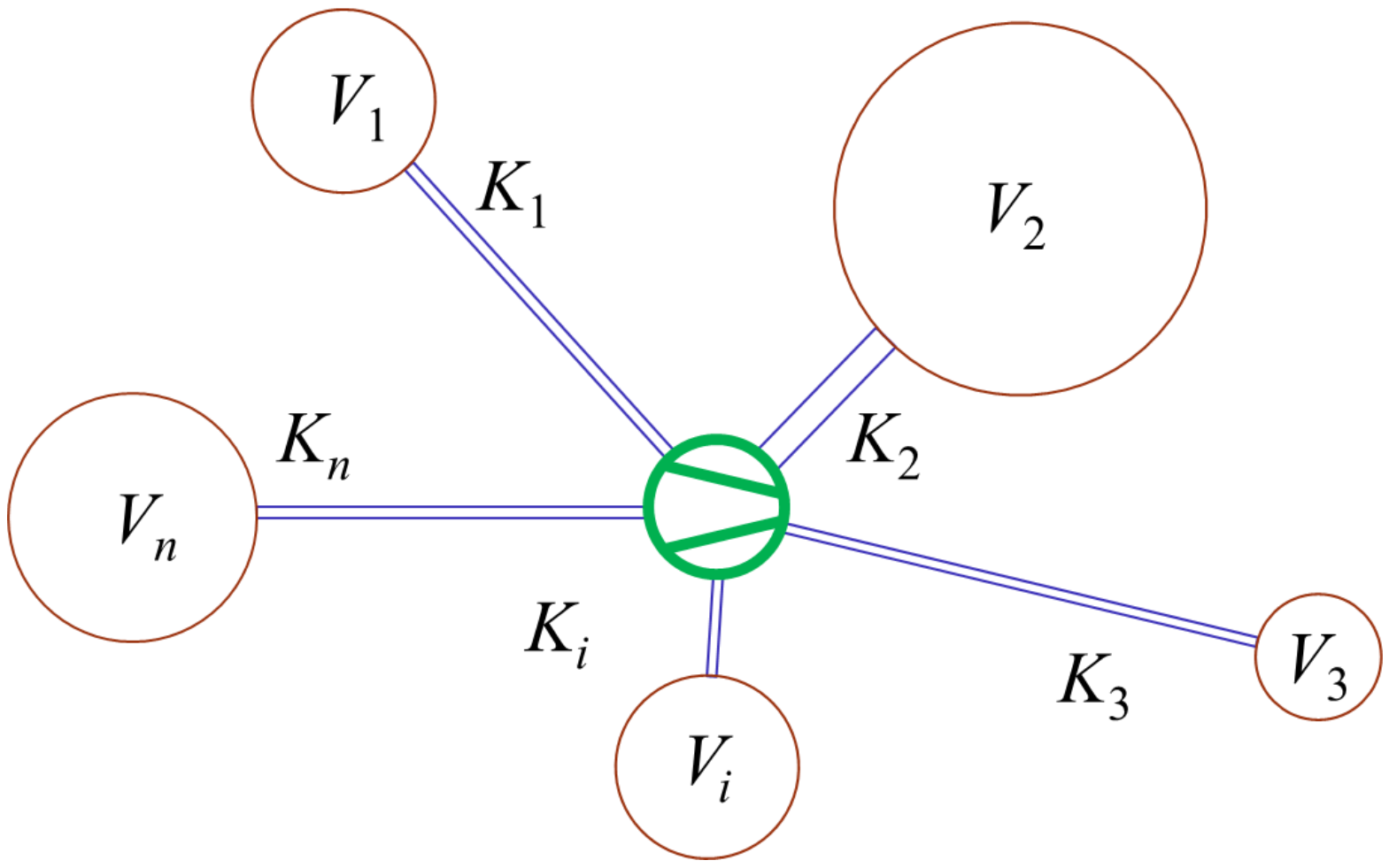

Figure 3: Schematics of a multi chamber vacuum system connected in parallel to a single pump.

Let $i$ and $j$, with $i \neq j$ be indexes representing the chambers. The pressure congruency in all chambers constrains the possible solutions for $\{K_i\}$. That is:

$$p_i(t) \equiv p_j(t) \Rightarrow \begin{cases} \dfrac{S_i}{V_i} = \dfrac{S_j}{V_j} = C_1 & (14) \\ p_{Ri} = p_{Rj} = C_2 \text{ .} & (15) \end{cases}$$

The total pump rate is the sum of the pump rates at each tube:

$$S_1 + S_2 + \ldots + S_i \ldots + S_n = S \text{ .} \tag{16}$$

Then, using (14) in (16), it follows that:

$$C_1 = \frac{S}{V_1 + V_2 + \ldots + V_i \ldots + V_n} = \frac{S}{V} \text{ ,} \tag{17}$$

and, replacing $C_1$ back in (14) we find:

$$S_i = \frac{V_i}{V} S \text{ .} \tag{18}$$

Similarly, (15) gives

$$\frac{V_i}{K_i} = \frac{V_j}{K_j} = C_2 \text{ .} \tag{19}$$

Using $V_1 + \ldots + V_n = V$, it follows that:

$$C_2 = \frac{V}{K_1 + \ldots + K_n} \tag{20}$$

and replacing $C_2$ back in (19), we obtain our goal expression:

$$K_i - (K_1 + ... + K_n)\frac{V_i}{V} = 0 \quad . \tag{21}$$

This equation represents a homogeneous system of $n$ equations. Hence, there are infinite sets $\{K_i\}$ satisfying (21).

To solve (21) univocally, one of the $K_i$ must be known using some criterion. For example: one may want $\Delta p$ < 100 Pa at all times; or one may want the pump time interval to be smaller than a given tolerance ($\Delta t < T_{tolerance}$); or more directly, $K_1$ can be assigned a value. Remember that $K = \pi R^4/(8\mu L)$, where $L$ is probably fixed by the positions of the chamber and the pump, then $R_1$ can be chosen freely to define $K_1$.

By any criterion chosen, once $K_1$ is known, (21) results in:

$$K_i = \frac{K_1}{V_1} V_i \quad . \tag{22}$$

Figure 4 exemplifies a typical situation with 5 chambers at different distances from a common vacuum pump. The figure indicates the volumes (in $m^3$), the distances, and the radius that satisfy the conditions of uniform pressure drop. The criterion that we chose to solve (21) univocally is that the evacuation time to 100 Pa is just 10 % larger than the time for infinite tube conductance, i.e., $\Delta t = 1.1 \Delta t_\infty = 1900$ s. In these conditions $p_R = 60.4$ Pa for all tubes, which allows the determination of $\{R_i\}$. For example, for chamber #1, $S_1 = 0.0016$ $m^3/s$, $K_1 = 2.65 \times 10^{-5}$ $m^3 s^{-1} Pa^{-1}$ implying $R_1 = 1.61$ cm, and so on.

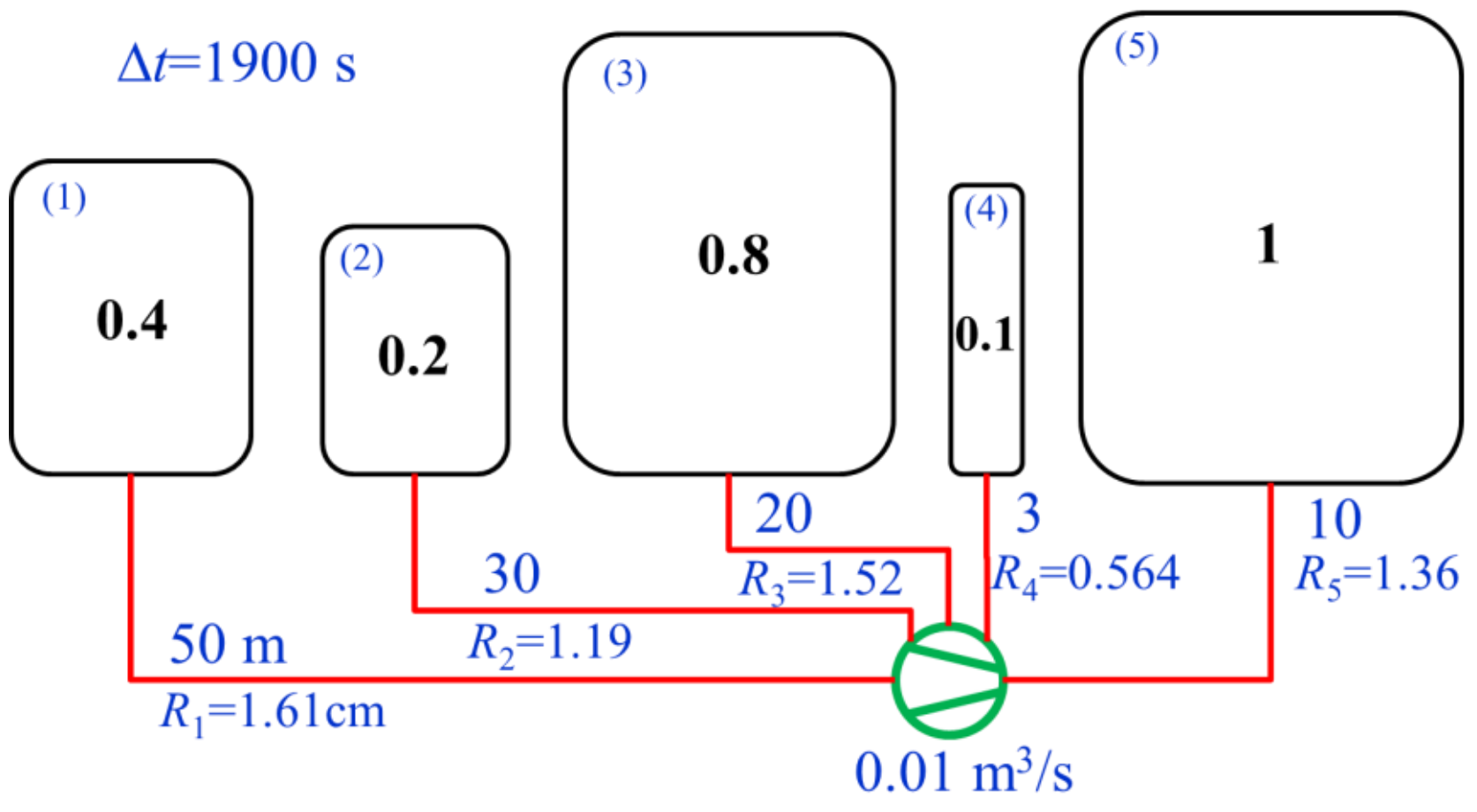

Figure 4: Given the positions of the chambers relative to the pump in addition to a specified evacuation time interval, one can determine the radii of the tubes that homogenize the pressure decay in all chambers as indicated.

These results can be useful even if customized tubes are not available to match precisely the conditions of homogeneous pressure decay. If $R_{ideal}$ is the ideal radius predicted form (21) and $\{R_{avail}\}$ is the set of tubes radii available, then the optimal radius to employ from the set obeys the condition $\min\left(\sqrt[4]{\{R_{avail}\}} - \sqrt[4]{R_{ideal}}\right)$. In other words, the optimal available tube will be the one that has the fourth root of the radius closest to the fourth root of the radius of the ideal tube.

## 5. Conclusions

The derivation of eq. (9) is one of our main results. Although it has been derived previously in the literature, one may find those references difficult to get. It predicts $p(t)$ given the geometrical parameters of the vacuum system. The pressure drop due to the finite conductance of the tube can be very different from the exponential decay as demonstrated with numerical examples.

The set $\{K_i\}$ we derive (eq. 22) enables one to determine the ideal radii of the tubes in the system, provided that one of the tube's radii is known, as discussed. We find eq. (21) remarkable in several aspects. Starting with its overall simplicity and symmetry, it is surprising that the $\{K_i\}$ is not univocal even though every geometrical and physical parameter in the system are defined. To make the set $\{K_i\}$ univocal, one of the $K_i$ must be fixed.

The volume rate $S_i$ (Eq. 18) is also interesting on its own. The simplicity of $S_i$ is particular to the conditions of pressure homogeneity. If the pressures in the chambers are allowed to be independent, then the $\{S_i\}$ has only numerical solution [5].

Even if experimentalists cannot make use of the ideal conditions predicted for the radii of the tubes, one can still benefit from this analysis by using radii that are the closest to our recommendations.

## AUTHOR DECLARATIONS

**Conflict of Interest**: The authors have no conflicts to disclose.

**Data Availability**: The data that support the findings of this study are available from the corresponding author upon request.

**Contributions**: F.F. Dall'Agnol: Conceptualization, Design, Writing, Modeling, Algebraic manipulation. F. Vieira: Review, Writing, Algebraic manipulation. F.T. Degasperi: Conceptualization, Review, Writing, Referencing.

**References**.

---